\documentclass[%
reprint,
amsmath,amssymb,
aps,
prb,
]{revtex4-2}

\usepackage{graphicx}% Include figure files
\usepackage{dcolumn}% Align table columns on decimal point
\usepackage{bm}% bold math
\usepackage{xcolor,MnSymbol,float}

\newcommand{\mybf}[1]{\boldsymbol{\mathbf{#1}}}

\begin{document}
	
	\preprint{APS/123-QED} 
	
	\title{Keldysh Theory of Thermal Transport in Multiband Hamiltonians}

	\author{Luqman Saleem}
	\email{luqman.saleem@kaust.edu.sa}
	\affiliation{%
		Physical Science and Engineering Division, King Abdullah University of Science and Technology (KAUST), Thuwal 23955, Saudi Arabia
	}

	\author{Udo Schwingenschlögl}
	\email{udo.schwingenschlogl@kaust.edu.sa}
	\affiliation{%
		Physical Science and Engineering Division, King Abdullah University of Science and Technology (KAUST), Thuwal 23955, Saudi Arabia
	}
	
	\author{Aurélien Manchon}
	\email{aurelien.manchon@univ-amu.fr}
	\affiliation{Aix-Marseille Universit\'e, CNRS, CINaM, Marseille, France
	}

	%\date{\today}
	
	\begin{abstract}
		We establish a comprehensive theoretical framework for systems subjected to a static uniform temperature gradient, employing the non-equilibrium Keldysh-Dyson formalism. This framework interprets the statistical force due to the temperature gradient as a mechanical force, utilizing both Luttinger's scalar and Moreno-Coleman-Tatara's vector potentials, which collectively emulate the gauge invariance stemming from the conservation of energy. Our approach has the ability to treat heat current and heat magnetization on an equal footing, thereby extending and generalizing previous formalisms. The derived result for the thermal conductivity is applied to investigate the thermal characteristics of Weyl magnons in a stacked honeycomb ferromagnet featuring a trivial insulator phase, a magnon Chern insulator phase, and three Weyl magnon phases. Against the expectation from the Berry curvature, the magnon Chern insulator phase exhibits the highest transverse thermal conductivity.
	\end{abstract}
	
	\maketitle
	
	%\tableofcontents
	
	\section{Introduction} \label{sec:level1}
	
	The theoretical description of quantum transport induced by an electromagnetic stimulus is well established in condensed matter physics \cite{Kubo1957I, Kubo1957II, Smrcka1977, RevModPhys.58.323, Onoda2006}. In the minimal coupling scheme, the electromagnetic field is inserted into the Hamiltonian via Peierl's substitution \cite{Peierls1933} and the adoption of an appropriate gauge (e.g., length, velocity, Landau) enables the study of a wealth of phenomena \cite{RevModPhys.82.1959, RevModPhys.82.1539, RevModPhys.87.1213, RevModPhys.95.011002}. In contrast, the theoretical description of quantum transport under a temperature gradient remains a challenging task due to the absence of a clear fundamental theory for this type of non-equilibrium force \cite{PhysRev.135.A1505,Chernodub2022}. This hinders the progress of emergent fields of research aiming at harvesting thermal transport, such as spin caloritronics \cite{Bauer2012,Boona2014,Yokoyama2014,Yu2017,UCHIDA2021}, topological magnonics \cite{PhysRevB.97.081106,PhysRevB.102.024409,Barman2021,Wang2021,McClarty2022,2305.14861}, quantum heat transport \cite{RevModPhys.93.041001,Ridley2022}, and topological heat transport \cite{Rivas2017,Chernodub2022}.
	
	Unlike electromagnetic fields, a temperature gradient gives rise to a statistical force that cannot be readily incorporated into the Hamiltonian. Luttinger \cite{PhysRev.135.A1505} introduced a ``gravitational" scalar potential which enables the study of thermally induced transport of electrons \cite{Smrcka1977,PhysRevB.31.7291,PhysRevB.80.115111,PhysRevLett.107.236601,PhysRevB.90.060201,PhysRevB.90.115116,PhysRevB.104.035150}, magnons \cite{PhysRevLett.106.197202, Matsumoto2011, Murakami2011, PhysRevB.84.184406, PhysRevB.89.054420, PhysRevB.89.134409, PhysRevB.90.024412, Murakami2017, PhysRevB.99.014427, PhysRevB.100.100401, PhysRevB.101.024427, PhysRevResearch.2.023065, PhysRevLett.128.117201}, and phonons \cite{PhysRevB.86.104305,bhalla2021optical}. However, this approach results in diverging thermal transport coefficients in the zero temperature limit for systems without time reversal symmetry \cite{Agarwalla2011,PhysRevLett.106.197202,PhysRevB.84.184406}. To address this issue, various attempts were made, including Ref.\ \cite{PhysRevLett.107.236601}'s proposal to subtract the divergenceless equilibrium magnetization currents from the thermal transport coefficients calculated using Luttinger's scalar potential. However, the scaling law employed by this approach lacks a microscopic justification and cannot account for interaction effects and disorder.
	
	Shitade \cite{Shitade2014} presented a gauge-invariant theory of gravity by imposing local space-time translation symmetry, based on scalar and vector gauge fields. The scalar gauge field was identified to correspond to Luttinger's scalar potential, while the vector gauge field gives rise to tortional electric and magnetic fields. The author utilized this formalism to compute the thermal conductivity as a response to the tortional electric field and the heat magnetization as a response to the tortional magnetic field. However, the assumption of local space-time translation symmetry was not explicitly addressed in the context of a temperature gradient. Relying on the Cartan and Keldysh approaches in curved spacetime, the formalism also often exhibits technical complexity and lacks intuitive clarity.
	
	Moreno and Coleman \cite{moreno1996thermal} represented a temperature gradient by a vector potential rather than a scalar potential. This approach was later adopted by Tatara \cite{PhysRevLett.114.196601,PhysRevB.92.064405} to address the issue of diverging thermal transport coefficients. It was argued that the vector potential, also known as the thermal vector potential, in conjunction with the magnetic vector potential that represents electromagnetic fields, can provide accurate thermal transport coefficients by considering diamagnetic heat currents. Nonetheless, the approach proved insufficient to fully address the issue of diverging thermal transport coefficients in systems without time reversal symmetry due to the distinct nature of thermal transport. The heat current, which quantifies entropy changes, must incorporate a component arising from the magnetization because the thermodynamic expression for heat inherently includes this component \cite{Obraz1,Obraz2,PhysRevB.55.2344,PhysRevB.80.214516}. The magnetization also influences electrically insulating excitations like phonons (lattice vibrations) and magnons (elementary magnetic excitations). Typically, such excitations possess non-zero orbital magnetic moments. In the presence of a temperature gradient these orbital magnetic moments induce magnetization currents, contributing to the physical currents. Consequently, an accurate description of thermal transport necessitates the consideration of both conventional heat currents, proportional to the temperature gradient, and magnetization currents, proportional to the magnetization gradient.
	
	The experimental discovery of a magnon-mediated Hall effect \cite{Onose2010} has spurred extensive research into understanding the magnon transport in the presence of a temperature gradient. Systems featuring Dzyaloshinskii-Moriya interaction \cite{Dzyaloshinsky1958,PhysRev.120.91} exhibit various intriguing transport phenomena, including the magnon spin-momentum locking \cite{PhysRevLett.119.107205}, magnon thermal Hall effect in both ferromagnetic  \cite{PhysRevLett.104.066403,PhysRevB.89.134409,PhysRevLett.106.197202,PhysRevB.84.184406} and antiferromagnetic \cite{PhysRevB.89.054420, PhysRevB.94.174444,PhysRevB.95.014422,PhysRevLett.121.097203,PhysRevB.98.094419,PhysRevLett.122.057204,PhysRevB.99.054409,PhysRevB.99.014427} materials, magnon spin Nernst effect \cite{PhysRevB.93.161106,PhysRevLett.117.217203,PhysRevLett.117.217202,PhysRevB.96.134425,PhysRevB.98.035424,PhysRevB.97.174407,PhysRevB.100.100401,PhysRevResearch.2.013079}, magnonic magnetization torque \cite{PhysRevB.90.224403,PhysRevB.93.161106,PhysRevB.95.165106}, and magnonic Edelstein effect \cite{PhysRevB.101.024427}. Additionally, topological phenomena were proposed, such as magnon topological insulators \cite{PhysRevB.87.144101,PhysRevB.89.134409,PhysRevB.90.024412,PhysRevB.96.224414,PhysRevB.98.060404,PhysRevB.97.081106,PhysRevB.99.224433,PhysRevLett.122.187203,PhysRevLett.125.217202} and Dirac \cite{PhysRevLett.127.217202}, Floquet \cite{Owerre2017,Owerre2019}, and Weyl \cite{PhysRevLett.117.157204,PhysRevB.97.094412,Owerre2018} magnons. Developing a thorough theoretical framework for describing the magnon transport in the presence of a temperature gradient is of utmost importance to unlock the potential of magnonics in next-generation spintronic devices \cite{Barman2021}.
	
	Unfortunately, most theoretical formulations of magnon transport rely on Luttinger's scalar potential. The widely used thermal conductivity formula of Matsumoto \textit{et al.} \cite{PhysRevLett.106.197202} attempts to address the divergence of the thermal transport coefficients by incorporating a correction factor resembling orbital magnetization, proportional to $\mybf{r}\times\mybf{v}$. However, this formula contains a position operator that is not well defined for extended systems. Furthermore, it primarily focuses on the contribution of the Berry curvature, neglecting other potential contributions, and is only applicable to non-interacting magnons. Given the recent surge of research on interacting magnons \cite{PhysRevX.11.021061,https://doi.org/10.48550/arxiv.2211.15157}, a comprehensive theory based on Green's functions to account for interaction effects is necessary.
	
	The Baym-Kadanoff \cite{Baym1961} or Keldysh-Dyson \cite{PhysRev.75.1736,keldysh1965diagram} formalism is a powerful tool for studying non-equilibrium systems. Its extension to temperature gradient perturbations will enable the calculation of both heat current and heat magnetization in both fermionic and bosonic systems. A Green's function formalism is also readily adaptable to multiband Hamiltonians and effects related to the Berry curvature of multiband Bloch states.  This will enable a more accurate description of the thermal transport in interacting and disordered systems, and pave the way for the design and optimization of magnonic devices.
	
	This study presents a theoretical framework to calculate linear and nonlinear responses to a steady and uniform temperature gradient. The temperature gradient is described by both Luttinger's scalar and thermal vector potentials, which together result in thermal electric and magnetic fields. Just as the charge current is proportional to the applied electric field, the heat current is proportional to the thermal electric field; and the heat magnetization is related to the thermal magnetic field. The response function is then computed within the Keldysh-Dyson formalism in the stationary case. 
	
	The present approach has major advantages over previous approaches \cite{PhysRevLett.107.236601,PhysRevLett.106.197202, Matsumoto2011, Murakami2011, PhysRevB.84.184406, PhysRevB.89.054420,Shitade2014,Murakami2017}. Firstly, the use of a vector potential to represent the temperature gradient eliminates previously neglected diamagnetic heat currents. Secondly, the heat currents can be expressed in terms of Fermi or Bose distribution functions, enabling the separation of intrinsic and extrinsic contributions to the thermal transport coefficients. Thirdly, the formalism is readily applicable to multiband systems, providing a versatile tool for investigating thermal transport in magnets with multiple magnetic sublattices as well as in magnetic heterostructures. Fourthly, the formalism can be extended to other transport quantities (e.g., particle transport, nonequilibrium densities) and time-dependent non-uniform problems. 
	
	This paper is structured as follows: Section \ref{sec:heat_and_TVP} introduces the thermal vector potential, its associated Hamiltonian, and the thermal electromagnetic field. Section \ref{sec:keldysh} develops a Keldysh-Dyson formalism for non-equilibrium Green's functions, expanding them in terms of the thermal electromagnetic field tensor and establishing a linear response theory. Section \ref{sec:heat_transport} derives the thermal conductivity, obtaining the heat current as a response to the thermal electric field and the heat magnetization as a response to the thermal magnetic field. It is shown that in the clean limit the result is equivalent to the formula of Matsumoto \textit{et al.} \cite{PhysRevLett.106.197202, Matsumoto2011, Murakami2011, PhysRevB.84.184406, PhysRevB.89.054420, Murakami2017}. Section \ref{sec:honeycomb_FM} focuses on the thermal properties of Weyl magnons in a honeycomb ferromagnet. Section \ref{sec:summary} provides a summary and discussion of the findings.
	
	\section{Heat Transport and Thermal Vector Potential} \label{sec:heat_and_TVP}
	The thermal vector potential $\mybf{A}_T(t,\mybf{r})$ is related to Luttinger's scalar potential $\phi(t,\mybf{r})$ and the temperature gradient as \cite{PhysRevLett.114.196601}
	\begin{align} \label{E:mechanical_fields}
		\partial_t\mybf{A}_T(t,\mybf{r})=\mybf{\nabla}\phi(t,\mybf{r})=\frac{\mybf{\nabla} T}{T},
	\end{align}
	with $\partial_t\equiv \partial/\partial t$. Unlike the electromagnetic theory, where the magnetic vector potential possesses U(1) gauge symmetry, $\mybf{A}_T(t,\mybf{r})$ does not possess gauge symmetry in the formal sense because the energy conservation arises from the global translation symmetry with respect to time. However, it can be shown that it couples to the energy in minimal form, i.e., $\mybf{p}\rightarrow\mybf{p}-\mybf{A}_T\epsilon_{ \mybf{k}}$, where $\mybf{p}$ is the canonical momentum and $\epsilon_{ \mybf{k}}$ is the energy in momentum space \cite{PhysRevLett.114.196601}. In real space-time operator form this relation becomes $\mybf{p}\rightarrow\mybf{p}-i\hbar\mybf{A}_T\partial_t$. The minimal form can be attributed to a ``gauge invariance'' resulting from the energy conservation law. In fact, one can attribute a part of the temperature gradient to Luttinger’s scalar potential and the other part to the thermal vector potential as
	\begin{equation}
		\frac{\mybf{\nabla} T}{T}=\mybf{\nabla}\phi(t,\mybf{r})+\partial_t\mybf{A}_T(t,\mybf{r}).
	\end{equation}
	Consequently, similar to the electromagnetic theory, we obtain gauge invariance under the transformations $\phi\to\phi+d\chi/dt$ and $\mybf{A}_T\to\mybf{A}_T-\mybf{\nabla}\chi$, where $\chi$ is a scalar field \cite{PhysRevLett.114.196601}. To elevate the thermal theory to the level of the electromagnetic theory, we define a thermal electric field 
	\begin{equation} \label{E:thermal_fields_E}
		\mybf{E}_T(t,\mybf{r})\equiv -\partial_t\mybf{A}_T(t,\mybf{r})-\mybf{\nabla}\phi(t,\mybf{r})
	\end{equation}
	and a thermal magnetic field
	\begin{equation} \label{E:thermal_fields_B}
		\mybf{B}_T(t,\mybf{r})\equiv \mybf{\nabla}\times\mybf{A}_T(t,\mybf{r}).
	\end{equation}
	$\mybf{E}_T(t,\mybf{r})$ gives rise to the mechanical force due to the temperature gradient. The heat current density can be defined as quantum expectation value 
	\begin{equation} \label{E:JQ0}
		\mybf{J}_Q(t,\mybf{r}) = \pm i \text{Tr}\big(\hat{\mybf{J}}_Q(t,\mybf{r})\hat{G}_{\mybf{E}_T}^<(t,\mybf{r})\big).
	\end{equation}
	Here and in the following, the upper sign is for Bosons and the lower sign is for Fermions. $\hat{\mybf{J}}_Q(t,\mybf{r})$ is the heat current density operator defined as product of the velocity operator and time derivative \cite{Shitade2014} and $\hat{G}_{\mybf{E}_T}^<(t,\mybf{r})$ is the lesser component of the Green's function (proportional to $\mybf{E}_T(t,\mybf{r})$). The hat sign indicates the matrix of operators in the internal degrees of freedom. Moreover, a thermodynamical definition can be constructed for the heat magnetization $\mybf{M}_Q$ by noticing that the thermal vector potential couples to the heat current density \cite{PhysRevB.102.235161},
	\begin{align} \label{E:MQ1}
		\mybf{M}_Q=-\left(\frac{\partial\Omega}{\partial\mybf{B}_T}\right)_{\mu,T}.
	\end{align}
	$\Omega = E - TS - \mu N$ is the free energy with $E$, $S$, $\mu$, and $N$ denoting the internal energy, entropy, chemical potential, and particle number, respectively.

	\section{Keldysh-Dyson formalism for the thermal vector potential} \label{sec:keldysh}
	We start by defining a general unperturbed Hamiltonian for Bosons or Fermions as
	\begin{align} \label{E:H_0}
		{\hat{H}}_0=\int d\mybf{r} {\Psi}^{\dagger}(\mybf{r})\left({\hat{h}}_0(\mybf{p})+\hat{V}(\mybf{r})\right)\Psi(\mybf{r}),
	\end{align}
	where ${\hat{h}}_0(\mybf{p})={\left(-i\hbar \mybf{\nabla}\right)}^2/2m$ is the non-interacting kinetic part, $\hat{V}(\mybf{r})$ is a general one-particle potential, and $\Psi^\dagger(\mybf{r}) $ and $\Psi(\mybf{r})$ are the creation and annihilation field operators, respectively.
	
	Moving forward, we adopt a four-vector notation: Time and space coordinates are represented as $x\equiv x^{\mu}=(t,\mybf{r})$ and $x_{\mu}=(-t,\mybf{r})$, derivatives are represented as ${\partial}_{x^{\mu}}=({\partial}_t, \mybf{\nabla})$ and ${\partial}_{x_{\mu}}=(-{\partial}_t,\mybf{\nabla})$, and potentials are represented as $A^{\mu}_T(x)=(\phi(x),{\mybf{A}}_{T}(x))$ and ${{(A_T)}_\mu}(x)=(-\phi(x),{ \mybf{A}}_{T}(x))$. We proceed by applying a temperature gradient via the thermal electromagnetic tensor 
	\begin{equation}
		F^{\mu\nu}_T(x)={\partial}_{x_{\mu}}A^{\nu}_T(x)-{\partial}_{x_{\nu}}A^{\mu}_T(x).
	\end{equation}
	 A ``gauge-invariant" Hamiltonian that incorporates both Luttinger's scalar and thermal vector potentials is formulated as
	\begin{align} 
		\label{E:H_t}
		\hat{H}(t) =& \int d\mybf{r}\Psi^{\dagger}(x)\left( \hat{h}_0(\mybf{p}-i\hbar\mybf{A}_T\partial_t) + \hat{V}(\mybf{r}) \right) \Psi(x) \nonumber\\
		& + i\hbar\int d\mybf{r}\phi(x)\Psi^{\dagger}(x)\partial_t \Psi(x).
	\end{align}
	The second term is Luttinger's Hamiltonian \cite{PhysRev.135.A1505}. While $\hat{H}(t)$ does not include interactions such as the particle-particle interaction, the following formalism still applies even when these interactions are present. However, a self-energy correction according to the Feynman rule is necessary.

	\subsection{Green's functions}
	The non-equilibrium Green's functions are defined on the Keldysh contour $C$ as \cite{RevModPhys.58.323}
	\begin{align} \label{E:G_C}
		\mathcal{G}(x_1,x_2)=-i \big\langle T_C{\left(\Psi(x_1)\Psi^\dagger(x_2)\right)}\big\rangle.
	\end{align}
	$T_C$ is the time-ordering operator. The Dyson equations are
	\begin{subequations} \label{E:DE1}
		\begin{align} 
			&\left(i\hbar D_{t_1}-{\hat{h}}_0(-i \hbar \mybf{D}_{x_1})\right)\mathcal{G}(x_1,x_2) \nonumber \\
			&-\int dx_3 \Sigma(x_1,x_3)\mathcal{G}(x_3,x_2)=\delta(x_1-x_2), \label{E:DE1a} \\
			&\mathcal{G}(x_1,x_2)\left(-i\hbar D_{t_2}-{\hat{h}}_0(i \hbar \mybf{D}_{x_2})\right) \nonumber \\
			&-\int dx_3\mathcal{G}(x_1,x_3)\Sigma(x_3,x_2)=\delta(x_1-x_2), \label{E:DE1b}
		\end{align}
	\end{subequations}
	where we define a gauge covariant derivative analogous to the electromagnetic theory as 
	\begin{equation}
		D_{{(x_n)}_\mu}=\partial_{{(x_n)}_\mu}+A_{T}^\mu(x_n)\partial_{t_n}.
	\end{equation}
	In real-time representation both the Green's function $\mathcal{G}$ and the self energy $\Sigma$ have three independent components,
	\begin{subequations} \label{E:G_Sigma_matrix}
		\begin{align} 
			\mathcal{G}(x_1,x_2)&=\left( \begin{array}{cc}
				\hat{G}^R(x_1,x_2) & 2\hat{G}^<(x_1,x_2) \\ 
				0 & \hat{G}^A(x_1,x_2) \end{array}
			\right),\label{E:G_Sigma_matrix_G}\\
			\Sigma(x_1,x_2)&=\left( \begin{array}{cc}
				\hat{\Sigma}^R(x_1,x_2) & 2\hat{\Sigma}^<(x_1,x_2) \\ 
				0 & \hat{\Sigma}^A(x_1,x_2) \end{array}
			\right).\label{E:G_Sigma_matrix_Sigma}
		\end{align}
	\end{subequations}
	The lesser, advanced, and retarded Green's functions are defined as
	\begin{align} \label{E:G_<RA}
		\begin{aligned}
			\hat{G}^<(x_1,x_2)&=\mp i\left\langle {\Psi}^\dagger(x_2){\Psi}(x_1)\right\rangle,\\
			\hat{G}^A(x_1,x_2)&=+i\theta(t_2-t_1)\left\langle {\left[{\Psi}(x_1),{\Psi}^\dagger(x_2)\right]}_{\mp }\right\rangle,\\
			\hat{G}^R(x_1,x_2)&=-i\theta(t_1-t_2)\left\langle {\left[{\Psi}(x_1),{\Psi}^\dagger(x_2)\right]}_{\mp}\right\rangle.
		\end{aligned}
	\end{align}
	The lesser component is directly connected to the density matrix and can be used to calculate expectation values of operators.
	
	In a compact form, the Dyson equations (\ref{E:DE1a}) and (\ref{E:DE1b}) read
	\begin{subequations} \label{E:DE2}
		\begin{align}
			\left(\left((\mathcal{G}^{(0)})^{-1}-\Sigma\right)\star \mathcal{G}\right)(x_1,x_2)&=\delta(x_1-x_2), \label{E:DE2a}\\
			\left(\mathcal{G}\star \left((\mathcal{G}^{(0)})^{-1}-\Sigma\right)\right)(x_1,x_2)&=\delta(x_1-x_2) \label{E:DE2b}
		\end{align}
	\end{subequations}
	with the convolution star product defined as
	\begin{align} \label{E:star_real}
		(\mathcal{A}\star\mathcal{B})(x_1,x_2)=\int dx_3\mathcal{A}(x_1,x_3)\mathcal{B}(x_3,x_2)
	\end{align}
	and
	\begin{align} \label{E:G_0-1_real}
		\begin{aligned}
			(\mathcal{G}^{(0)})^{-1}(x_1,x_2)=\left(i\hbar D_{t_1}-{\hat{h}}_0(-i \hbar \mybf{D}_{x_1})\right) \delta(x_1-x_2),\\
			=\left(-i\hbar D_{t_2}-{\hat{h}}_0(i\hbar\mybf{D}_{x_2})\right)\delta(x_1-x_2).
		\end{aligned}
	\end{align}
	
	We transform the $(x_1,x_2)$ coordinates into the relative microscopic $x$ and center of mass macroscopic $X\equiv X^\mu$ coordinates using the Wigner representation $(X,x)=\left(1/2(x_1+x_2),x_1-x_2\right)$. We continue to the $(X,p)$ coordinates by Fourier transforming $x^\mu$ and $x_\mu$ to $p^\mu\equiv(\epsilon,\mybf{p})$ and $p_\mu\equiv(-\epsilon,\mybf{p})$, respectively. It is known that in the Wigner representation the convolution star product becomes the Moyal product \cite{Haug2008}
	\begin{align} \label{E:star_momentum}
		(\mathcal{A}\star\mathcal{B})(X,p)=\mathcal{A}(X,p)\text{e}^{i\hbar /2\left(\partial_{X^{\mu}}\partial_{p_{\mu}}-\partial_{p_{\mu}}\partial_{X^{\mu}}\right)}\mathcal{B}(X,p).
	\end{align} 
	To write the Dyson equations in the $(X,p)$ coordinates, we calculate $(\mathcal{G}^{(0)})^{-1}(X,p)$ by defining
	\begin{align} \label{E:pi_mu_real}
		\pi_\mu(x_1,x_2)=-i\hbar D_{x_1^\mu}\delta(x_1-x_2)=i\hbar D_{x_2^\mu}\delta(x_1-x_2)
	\end{align} 
	and Fourier transforming
	\begin{align} \label{E:pi_mu_momentum}
		\pi_\mu(X,p)&=\int dx e^{-ip_\mu x^\mu/\hbar}  \pi_\mu(X,x)\nonumber\\
		&=p_\mu-\epsilon(A_{T})_\mu(X,0) + i\hbar  \left(\partial_t(A_{T})_\mu(X,t)\right)_{t=0}.
	\end{align}
	The last term is neglected, as we are interested in the behavior of the system for $t>0$. $\pi_\mu(X,p)$ is the mechanical momentum (which is different from the canonical momentum $p_\mu$).
	
	Finally, we have in $(X,p)$ coordinates 
	\begin{align} \label{E:G_0-1_momentum}
		(\mathcal{G}^{(0)})^{-1}(\pi(X,p))={\pi}^0-{\hat{h}^*}_{0}( \mybf{\pi}),
	\end{align}
	where ${\hat{h}^*}_{0}(\mybf{\pi})$ is the generalization of ${\hat{h}}_{0}(\mybf{p})$ in the presence of Luttinger's scalar and thermal vector potentials. It is obtained by replacing all direct products by star products \cite{Onoda2006}. The final form of the Dyson equations in terms of the Moyal product is
	\begin{subequations} \label{E:DE3}
		\begin{align}
			{\left(\left((\mathcal{G}^{(0)})^{-1}-\Sigma\right)\star \mathcal{G}\right)}(X,p)&=1,\label{E:DE3a}\\
			{\left(\mathcal{G}\star \left((\mathcal{G}^{(0)})^{-1}-\Sigma\right)\right)}(X,p)&=1. \label{E:DE3b}
		\end{align}
	\end{subequations}
	 
	\vspace{0.5pt}
	
	\subsection{Static uniform thermal field}
	The off-diagonal elements of $F_{T}^{\mu\nu}$ are the components of the thermal electric and magnetic fields. We decompose $A_{T}^{\mu}(X,0)$ into ${\tilde{A}}^{\mu}_T(X,0)$, which covers the static uniform parts of the thermal electric and magnetic fields (${\tilde{F}}^{\mu\nu}_T={\partial}_{X_{\mu}}{\tilde{A}}^{\nu}_T-{\partial}_{X_{\nu}}{\tilde{A}}^{\mu}_T=\text{constant}$) and $ \delta A^{\mu}_T(X,0)=A^{\mu}_T(X,0)-{\tilde{A}}^{\mu}_T(X,0)$, which covers the dynamic and the non-uniform parts. Similarly, the static uniform mechanical momentum is ${\tilde{\pi}}^{\mu}(X,p)=p^\mu-\epsilon\tilde{A}_{T}^\mu(X,0)$. Changing the variables of the Moyal product (\ref{E:star_momentum}) and Dyson equations (\ref{E:DE3a}) and (\ref{E:DE3b}) from the canonical momentum $p$ to the static uniform kinetic momentum $\tilde{\pi}$ yields
	\begin{widetext}
		\begin{subequations} \label{E:DE4}
			\begin{align}
				{\left(\mathcal{A}\star\mathcal{B}\right)}(X,\tilde{\pi})=\mathcal{A}(X,\tilde{\pi}){\text{e}^{i\hbar/2\left(\partial_{X^{\mu}}\partial_{{\tilde{\pi}}_{\mu}}-\partial_{{\tilde{\pi}}_{\mu}}\partial_{X^{\mu}}+{\tilde{\pi}}^0{\tilde{F}}^{\mu\nu}_T\partial_{{\tilde{\pi}}^{\mu}}\partial_{{\tilde{\pi}}^{\nu}}\right)}}\mathcal{B}(X,\tilde{\pi}), \label{E:DE4a} \\
				\left({\tilde{\pi}}^0-\epsilon\delta\phi(X,0)-{\hat{h}}_{0^*}(\tilde{\mybf{\pi}}-\epsilon\delta{\mybf{A}}_T(X,0))-\Sigma(X,\tilde{\pi})\right)\star \mathcal{G}(X,\tilde{\pi})=1,\label{E:DE4b}\\
				\mathcal{G}(X,\tilde{\pi})\star\left({\tilde{\pi} }^0-\epsilon\delta\phi(X,0)-{\hat{h}}_{0^*}(\tilde{\mybf{\pi}} -\epsilon \delta {\mybf{A}}_T(X,0))-\Sigma(X,\tilde{\pi})\right)=1.\label{E:DE4c}
			\end{align}
		\end{subequations}
	\end{widetext}
	
	We address in the following only static uniform thermal electromagnetic fields, i.e., $\delta A_T(X,0)=0$, meaning that we do not need to distinguish between $\pi$ and $\tilde{\pi}$. Furthermore, we restrict ourselves to steady-state thermal electromagnetic fields, which simplifies (\ref{E:DE4a}), (\ref{E:DE4b}), and (\ref{E:DE4c}) to
	\begin{subequations} \label{E:DE5}
		\begin{align}
			(\mathcal{A}\star\mathcal{B})(\pi)=\mathcal{A}(\pi){\text{e}^{(i\hbar /2){\pi}^0F^{\mu\nu}_T\partial_{{\pi}^{\mu}}\partial_{{\pi}^{\nu}}}}&\mathcal{B}(\pi),\label{E:DE5a}\\
			\left({\pi}^0-{\hat{h}}_0(\mybf{\pi})-\Sigma(\pi)\right)\star \mathcal{G}(\pi)&=1,\label{E:DE5b}\\
			\mathcal{G}(\pi)\star\left({\pi}^0-{\hat{h}}_0(\mybf{\pi})-\Sigma(\pi)\right)&=1.\label{E:DE5c}
		\end{align}
	\end{subequations} 
	From now on, all quantities are functions of $\pi$ unless otherwise stated. 
	\subsection{Expansion of the Dyson equations}
	While often the Dyson equations (\ref{E:DE5b}) and (\ref{E:DE5c}) cannot be solved exactly, we can perturbatively expand them using the Moyal product (\ref{E:DE5a}), applying the Green's function ${\mathcal{G}}_0={({\pi}^0-{\hat{h}}_0(\mybf{\pi})-{\Sigma}_0)}^{-1}$ from the left (right) to the first (second) equation, and using the identity ${\mathcal{G}}^{-1}_0{\mathcal{G}}_0={\mathcal{G}}_0{\mathcal{G}}^{-1}_0=1$,
	\begin{widetext}
		\begin{subequations} \label{E:expansion2}
			\begin{align}
				\mathcal{G}&={\mathcal{G}}_0\left(1+(\Sigma-{\Sigma}_0)\mathcal{G}-\sum^{\infty }_{n=1}\frac{1}{n!}\left({\pi}^0-{\hat{h}}_0(\mybf{\pi})-\Sigma\right)\left(\prod^n_{i=1}\frac{i\hbar {\pi}^0}{2}F^{{\mu}_i{\nu}_i}_T\partial_{{\pi}^{{\mu}_i}}\partial_{{\pi}^{{\nu}_i}}\right)\mathcal{G}\right),\label{E:expansion2a}\\
				\mathcal{G}&=\left(1+\mathcal{G}(\Sigma-{\Sigma}_0)-\sum^{\infty }_{n=1}\frac{1}{n!}\mathcal{G}\left(\prod^n_{i=1}\frac{i\hbar {\pi}^0}{2}F^{{\mu}_i{\nu}_i}_T\partial_{{\pi}^{{\mu}_i}}\partial_{{\pi}^{{\nu}_i}}\right)\left({\pi}^0-{\hat{h}}_0(\mybf{\pi})-\Sigma\right)\right){\mathcal{G}}_0.\label{E:expansion2b}
			\end{align}
		\end{subequations}
	\end{widetext}
	Here and in the following, the subscript $0$ in the Green's functions and self-energies indicates that the quantity is taken in the absence of external fields. To expand Eqs. (\ref{E:expansion2a}) and (\ref{E:expansion2b}) in powers of $\hbar F^{\mu\nu}_T/2$ we aim for solutions of the form
	\begin{subequations} \label{E:solution1}
		\begin{align}
			\mathcal{G}&={\mathcal{G}}_0+\sum^{\infty }_{n=1}\frac{1}{n!}\left(\prod^n_{i=1}\frac{\hbar }{2}F^{{\mu}_i{\nu}_i}_T\right){\mathcal{G}}_{{\mu}_1{\nu}_1,\cdots,{\mu}_n{\nu}_n},\label{E:solution1a}\\ 
			\Sigma&={\Sigma}_0+\sum^{\infty }_{n=1}\frac{1}{n!}\left(\prod^n_{i=1}\frac{\hbar }{2}F^{{\mu}_i{\nu}_i}_T\right){\Sigma}_{{\mu}_1{\nu}_1,\cdots,{\mu}_n{\nu}_n}.\label{E:solution1b}
		\end{align}
	\end{subequations} 
	We find ${\mathcal{G}}_{{\mu}_1{\nu}_1,\cdots,{\mu}_n{\nu}_n}$ by inserting these equations into Eq.\ (\ref{E:expansion2a}) or (\ref{E:expansion2b}) and comparing the left and right hand sides. We can calculate by Eq.\ (\ref{E:G_Sigma_matrix}) the retarded, advanced, and lesser Green's functions up to any power of $\hbar F^{\mu\nu}_T/2$. Below, we explicitly give the zeroth order (equilibrium Green's function theory) and first order (linear response theory) results.
	
	The equilibrium Green's function theory is derived from the equilibrium Dyson equation 
	\begin{equation}
		{\left({\epsilon }-{\hat{h}}_0(\mybf{p})-{\Sigma}_0(p)\right)}{\mathcal{G}}_0(p)=1
	\end{equation} 
	as
	\begin{subequations} \label{E:G_zero_order}
		\begin{align}
			G^{R(A)}_0(p)&={\left(\epsilon-{\hat{h}}_0(\mybf{p})-{\Sigma}^{R(A)}_0(p)\right)}^{-1},\label{E:G_zero_order_a}\\
			G^<_0(p)&=\mp\left(G^A_0(p)-G^R_0(p)\right)f_{\mp },\label{E:G_zero_order_b}\\
			{\Sigma}^<_0(p)&=\mp\left({\Sigma}^A_0(p)-{\Sigma}^R_0(p)\right)f_{\mp },\label{E:G_zero_order_c}
		\end{align}
	\end{subequations} 
	where $f_{\mp}(\epsilon)={(e^{\beta\epsilon}\mp 1)}^{-1}$ denotes the Bose and Fermi distribution functions.

	The linear response theory is derived by linearizing Eq.\ (\ref{E:expansion2a}) or (\ref{E:expansion2b}) in $\hbar F^{\mu\nu}_T/2$,
	\begin{align} \label{E:G_munu1}
		{\mathcal{G}}_{\mu\nu}={\mathcal{G}}_0{\Sigma}_{\mu\nu}{\mathcal{G}}_0+i{\pi}^0{\mathcal{G}}_0\left({\partial}_{{\pi}^{\mu}}{\mathcal{G}}^{-1}_0\right){\mathcal{G}}_0\left(\partial_{{\pi}^{\nu}}{\mathcal{G}}^{-1}_0\right){\mathcal{G}}_0.
	\end{align} 
	${\mathcal{G}}_{\mu\nu}$ is antisymmetric under the exchange of $\mu $ and $\nu$, because $F^{\mu\nu}_T$ is antisymmetric under this exchange. By rewriting Eq.\ (\ref{E:G_munu1}) as
	\begin{align} \label{E:G_munu2}
		{\mathcal{G}}_{\mu\nu}&={\mathcal{G}}_0{\Sigma}_{\mu\nu}{\mathcal{G}}_0 +\frac{i{\pi}^0}{2}\mathcal{G}_0 \left({\partial}_{{\pi}^{\mu}}{\mathcal{G}}^{-1}_0\right){\mathcal{G}}_0\left({\partial}_{{\pi}^{\nu}}{\mathcal{G}}^{-1}_0\right) \mathcal{G}_0 \nonumber \\
		&-\frac{i{\pi}^0}{2}\mathcal{G}_0\left({\partial}_{{\pi}^{\nu}}{\mathcal{G}}^{-1}_0\right){\mathcal{G}}_0\left({\partial}_{{\pi}^{\mu}}{\mathcal{G}}^{-1}_0\right)\mathcal{G}_0
	\end{align}
	we obtain the thermal electric and magnetic Green's functions
	\begin{subequations} \label{E:G_ETBT}
		\begin{align}
			{\mathcal{G}}_{{ \mybf{E}}_T}&={\mathcal{G}}_0{\Sigma}_{{ \mybf{E}}_T}{\mathcal{G}}_0
			+\frac{i{\pi}^0}{2}\mathcal{G}_0\left({\partial}_{{\pi}^0}{\mathcal{G}}^{-1}_0\right){\mathcal{G}}_0\left(\mybf{\nabla_\pi}{\mathcal{G}}^{-1}_0\right)\mathcal{G}_0 \nonumber \\
			&-\frac{i{\pi}^0}{2}\mathcal{G}_0\left(\mybf{\nabla_\pi}{\mathcal{G}}^{-1}_0\right){\mathcal{G}}_0\left({\partial}_{{\pi}^0}{\mathcal{G}}^{-1}_0\right)\mathcal{G}_0,\label{E:G_ET}\\
			{\mathcal{G}}_{{ \mybf{B}}_T}&={\mathcal{G}}_0\left({\Sigma}_{{ \mybf{B}}_T}+\frac{i{\pi}^0}{2}\left(\mybf{\nabla_\pi}{\mathcal{G}}^{-1}_0\times{\mathcal{G}}_0\left(\mybf{\nabla_\pi}{\mathcal{G}}^{-1}_0\right)\right)\right){\mathcal{G}}_0.\label{E:G_BT}
		\end{align}
	\end{subequations}
	
	We can decomposed these matrix Green's functions into their components using the fact that for a product $\mathcal{G}={\mathcal{G}}_1{\mathcal{G}}_2\cdots {\mathcal{G}}_n$ of matrix Green's function the retarded, advanced, and lesser components are
	\begin{align} \label{E:GRAG<_identity}
		\hat{G}^{R(A)}&=\hat{G}^{R(A)}_1\hat{G}^{R(A)}_2\cdots \hat{G}^{R(A)}_n,\\
		\hat{G}^<&=\hat{G}^<_1\hat{G}^A_2\cdots \hat{G}^A_n+\hat{G}^R_1\hat{G}^<_2\cdots \hat{G}^A_n+\cdots+ \hat{G}^R_1\hat{G}^R_2\cdots \hat{G}^<_n. \nonumber
	\end{align}
	Employing the relations
	\begin{align}
		\begin{split}
				\mybf{\nabla_{\pi}}\hat{G}^R_0=&-\hat{G}^R_0\mybf{\nabla_{\pi}}({\hat{G}_0}^{R})^{-1}\hat{G}^R_0, \\
			\partial_{\pi^0}\hat{G}^R_0=&-\hat{G}^R_0\partial_{\pi^0}({\hat{G}_0}^{R})^{-1}\hat{G}^R_0,
		\end{split}
	\end{align}
	we obtain
	\begin{widetext}
		\begin{subequations} \label{E:GRA<_ETBT}
			\begin{align}
				\hat{G}^{R(A)}_{{ \mybf{E}}_T}=&\ \hat{G}^{R(A)}_0\left({\hat{\Sigma}}^{R(A)}_{{ \mybf{E}}_T}+\frac{i{\pi}^0}{2}\left({\partial}_{{\pi}^0}
				\left({\hat{G}_0}^{R(A)}\right)^{-1}
				\hat{G}^{R(A)}_0\mybf{\nabla_\pi}
				\left({\hat{G}_0}^{R(A)}\right)^{-1}
				-\mybf{\nabla_\pi}\left({\hat{G}_0}^{R(A)}\right)^{-1}
				\hat{G}^{R(A)}_0{\partial}_{{\pi}^0}\left({\hat{G}_0}^{R(A)}\right)^{-1}\right)\right)\hat{G}^{R(A)}_0,\label{E:GRA_ET}\\
				\hat{G}^{R(A)}_{{ \mybf{B}}_T}=&\ \hat{G}^{R(A)}_0\left({\hat{\Sigma}}^{R(A)}_{{ \mybf{B}}_T}+\frac{i{\pi}^0}{2}\left(\mybf{\nabla_\pi}\left({\hat{G}_0}^{R(A)}\right)^{-1}\times \hat{G}^{R(A)}_0\mybf{\nabla_\pi}\left({\hat{G}_0}^{R(A)}\right)^{-1}\right)\right)\hat{G}^{R(A)}_0,\label{E:GRA_BT}\\
				\hat{G}^<_{{ \mybf{E}}_T}=&\mp\left(\hat{G}^A_{{ \mybf{E}}_T}-\hat{G}^R_{{ \mybf{E}}_T}\right)f_{\mp }+\hat{G}^R_0\left({\hat{\Sigma}}^<_{ \mybf{E}_T}\pm\left({\hat{\Sigma}}^A_{ \mybf{E}_T}-{\hat{\Sigma}}^R_{ \mybf{E}_T}\right)f_{\mp }\right)\hat{G}^A_0\nonumber\\
				&\pm\frac{i{\pi}^0}{2}\hat{G}^R_0\left(\left({\hat{\Sigma}}^A_0-{\hat{\Sigma}}^R_0\right)\hat{G}^A_0\mybf{\nabla_\pi}\left({\hat{G}_0}^A\right)^{-1}-\mybf{\nabla_\pi}\left({\hat{G}_0}^{R}\right)^{-1}\hat{G}^R_0\left({\hat{\Sigma}}^A_0-{\hat{\Sigma}}^R_0\right)\right)\hat{G}^A_0f_{\mp }',\label{E:G<_ET}\\
				\hat{G}^<_{{ \mybf{B}}_T}=&\mp\left(\hat{G}^A_{{ \mybf{B}}_T}-\hat{G}^R_{{ \mybf{B}}_T}\right)f_{\mp }+\hat{G}^R_0\left({\hat{\Sigma}}^<_{{ \mybf{B}}_T}\pm\left({\hat{\Sigma}}^A_{{ \mybf{B}}_T}-{\hat{\Sigma}}^R_{{ \mybf{B}}_T}\right)f_{\mp }\right)\hat{G}^A_0 \label{E:G<_BT}
			\end{align}
		\end{subequations}
	\end{widetext} 
	with the abbreviations $f_\mp \equiv f_\mp(\pi^0)$ and $f_{\mp }' \equiv {\partial}_{{\pi}^0}f_{\mp }$. These equations have solutions of the form
	\begin{subequations} \label{E:G_first_order_sol}
		\begin{align}
			\hat{G}^<_{{ \mybf{E}}_T}&=\hat{G}^<_{{ \mybf{E}}_T,\text{I}}f_{\mp}+\hat{G}^<_{{ \mybf{E}}_T,\text{II}}f_{\mp}',\label{E:G_first_order_sola}\\
			{\hat{\Sigma}}^<_{{ \mybf{E}}_T}&={\hat{\Sigma}}^<_{{ \mybf{E}}_T,\text{I}}f_{\mp}+{\hat{\Sigma}}^<_{{ \mybf{E}}_T,\text{II}}f_{\mp}',\label{E:G_first_order_solb}\\
			\hat{G}^<_{{ \mybf{B}}_T}&=\hat{G}^<_{{\mybf{B}}_T,\text{I}}f_{\mp}+\hat{G}^<_{{ \mybf{B}}_T,\text{II}}f_{\mp}',\label{E:G_first_order_solc}\\
			{\hat{\Sigma}}^<_{{ \mybf{B}}_T}&={\hat{\Sigma}}^<_{{\mybf{B}}_T,\text{I}}f_{\mp}+{\hat{\Sigma}}^<_{{ \mybf{B}}_T,\text{II}}f_{\mp}'\label{E:G_first_order_sold}.
		\end{align}
	\end{subequations}
	The first and second terms always represent the interband and intraband contributions, respectively \cite{Onoda2006}. By inserting the solutions into Eqs. (\ref{E:GRA_ET})-(\ref{E:G<_BT}) and comparing the left and right hand sides, we find
	\begin{subequations} \label{E:G_first_order}
		\begin{align}
			\hat{G}^<_{ \mybf{E},\text{I}}=&\mp\left(\hat{G}^A_{{ \mybf{E}}_T}-\hat{G}^R_{{ \mybf{E}}_T}\right), \label{E:G_first_ordera}\\
			\hat{G}^<_{{ \mybf{E}}_T,\text{II}}=&\ \hat{G}^R_0{\hat{\Sigma}}^<_{{ \mybf{E}}_T,\text{II}}\hat{G}^A_0 \mp\frac{i{\pi}^0}{2}\mybf{\nabla_\pi}(\hat{G}^R_0+\hat{G}^A_0) \nonumber \\
			&\pm i{\pi}^0\hat{G}^R_0\mybf{\nabla_\pi}({\hat{h}}_0(\mybf{\pi})+\frac{1}{2}\left({\hat{\Sigma}}^R_0+{\hat{\Sigma}}^A_0\right))\hat{G}^A_0, \label{E:G_first_orderb}\\
			{\hat{\Sigma}}^<_{ \mybf{E},\text{I}}=&\mp\left({\hat{\Sigma}}^A_{{ \mybf{E}}_T}-{\hat{\Sigma}}^R_{{ \mybf{E}}_T}\right),\label{E:G_first_orderc}\\
			\hat{G}^<_{{ \mybf{B}}_T,\text{I}}=&\mp\left(\hat{G}^A_{{ \mybf{B}}_T}-\hat{G}^R_{{ \mybf{B}}_T}\right),\label{E:G_first_orderd}\\
			\hat{G}^<_{{ \mybf{B}}_T,\text{II}}=&\ \hat{\Sigma}^<_{{\mybf{B}}_T,\text{II}}=0 \label{E:G_first_ordere},\\
			{\hat{\Sigma}}^<_{{ \mybf{B}}_T,\text{I}}=&\mp\left({\hat{\Sigma}}^A_{{ \mybf{B}}_T}-{\hat{\Sigma}}^R_{{ \mybf{B}}_T}\right).\label{E:G_first_orderf}
		\end{align}
	\end{subequations}
	Eqs. (\ref{E:GRA_ET}), (\ref{E:GRA_BT}), (\ref{E:G_first_order_sola})-(\ref{E:G_first_order_sold}), and (\ref{E:G_first_ordera})-(\ref{E:G_first_orderf}) provide a complete description. 
	
	\section{Heat Transport} \label{sec:heat_transport}
	The thermal conductivity tensor is defined as \cite{PhysRevLett.107.236601}
	\begin{equation}
		\kappa_{ij}=\tilde{\kappa}_{ij}+\frac{2}{T}\epsilon_{ijk}M_Q^{k},
	\end{equation}
	where $\tilde{\kappa}_{ij}$ is the contribution from heat current density, $M_Q^{k}$ is the $k$th component of $\mybf{M}_Q$, and $\epsilon_{ijk}$ is the Levi-Civita tensor. In the Wigner representation Eq.\ (\ref{E:JQ0}) becomes \cite{Shitade2014}
	\begin{align}
		\mybf{J}_Q&=\pm i \text{Tr}\int\frac{d\pi^0}{2\pi}(\pi^0\mybf{\hat{v}}\mathcal{G})^<,\nonumber\\
		&=\pm i\hbar \text{Tr}\int\frac{d\pi^0}{2\pi}\pi^0
		\mybf{\hat{v}}\left(\hat{G}_{\mybf{E}_T,\text{I}}^< f_\mp+\hat{G}_{\mybf{E}_T,\text{II}}^< f'_\mp\right)\mybf{E}_T.\label{E:JQ2}
	\end{align}
	Its contribution to ${\kappa}_{ij}$ is
	\begin{align} \label{E:tildek_ij1}
		\tilde{\kappa}_{ij}&=\pm\frac{i\hbar}{T} \text{Tr}\int\frac{d\pi^0}{2\pi}\pi^0
		\hat{v}_i\left(\hat{G}_{\mybf{E}_T,\text{I}}^< f_\mp+\hat{G}_{\mybf{E}_T,\text{II}}^< f'_\mp\right)_j,
	\end{align}
	where $\hat{v}_i$ is the velocity operator in direction $i$, the temperature gradient points in direction $j$, and $i,j=x,y,z$. Before calculating the contribution of the heat magnetization to $\kappa_{ij}$ by Eq.\ (\ref{E:MQ1}), it is common practice \cite{PhysRevLett.107.236601,Shitade2014} to calculate the auxiliary heat magnetization $\tilde{\mybf{M}}_Q=-\left(\partial K/\partial\mybf{B}_T\right)_{T,\mu}$, where $K=E-\mu N$ is the grand canonical energy. We obtain ${\tilde{ \mybf{M}}}_Q=\beta^{-1} \partial\left({\beta}^2{\mybf{M}}_Q\right)/\partial\beta$ and have in the Wigner representation
	\begin{align} \label{E:K}
		K=\pm i\text{Tr}\int\frac{d\pi^0}{2\pi}(\pi^0\mathcal{G})^<,
	\end{align} 
	resulting in
	\begin{align} \label{E:tildeMQ}
		\tilde{\mybf{M}}_Q &= \pm i\hbar\text{Tr}\int\frac{d\pi^0}{2\pi}\pi^0 \hat{G}_{\mybf{B}_T,\text{I}}^<f_{\mp}.
	\end{align}
	By integration with respect to $\beta$, we arrive at 
	\begin{align} \label{E:MQ2}
		\mybf{M}_Q &= \pm \frac{i\hbar}{\beta^2}\text{Tr}\int_{\infty}^{\beta}d\beta\int\frac{d\pi^0}{2\pi}\pi^0 \hat{G}_{\mybf{B}_T,\text{I}}^< \beta f_{\mp}
	\end{align}
	 and
	\begin{align} 
		\label{E:k_ij1}
		\kappa_{ij} =& \pm\frac{i\hbar}{T} \text{Tr}\int\frac{d\pi^0}{2\pi}\pi^0 \bigg(
			\hat{v}_i\left(\hat{G}_{\mybf{E}_T,\text{I}}^<\right)_j f_\mp + \hat{v}_i\left(\hat{G}_{\mybf{E}_T,\text{II}}^<\right)_j f'_\mp\nonumber\\
			&+\epsilon_{ijk}\frac{2}{\beta^2}\int_{\infty}^{\beta}d\beta\left(\hat{G}_{\mybf{B}_T,\text{I}}^<\right)_k \beta f_{\mp}
		\bigg).
	\end{align}
	Together with Eqs. (\ref{E:GRA<_ETBT}) and (\ref{E:G_first_order}) this equation provides an accurate description of the thermal transport properties in multiband systems and is the central result of this work. 
	
	\begin{figure}[b]
		\includegraphics[width=\linewidth]{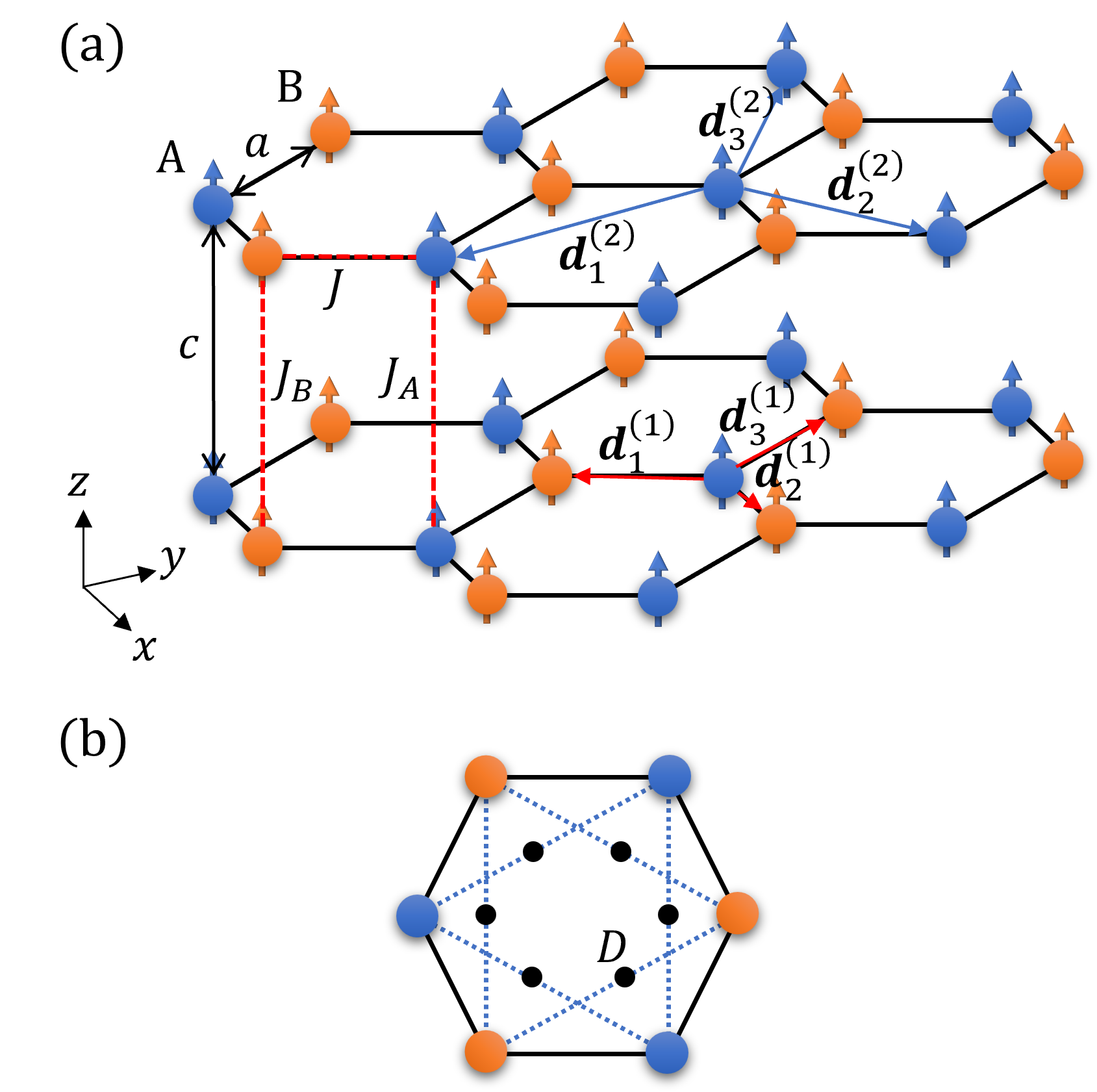}
		\caption{(a) Lattice representation of the stacked honeycomb ferromagnet with the spins oriented along the $z$ direction. Within a layer the nearest and next-nearest neighbors are separated by the vectors $\mybf{d}_\beta^{(1)}$ and $\mybf{d}_\beta^{(2)}$, respectively. The distance between two honeycomb lattice sites A and B is $a$ and the distance between adjacent layers is $c$. We set $a=c=1$. The intralayer Heisenberg interaction between nearest neighbors is $J$. The interlayer Heisenberg interaction between sites A and B in adjacent layers is $J_A$ and $J_B$, respectively.
		(b) The Dzyaloshinskii-Moriya interaction $D$, represented by black dots, acts perpendicularly to the layers and involves next-nearest neighbors.}
		\label{fig:honeycomb}
	\end{figure}
	
	To prove that our approach is consistent with the literature we consider the clean limit $\hat{\Sigma}=0$. We have $ \hat{G}_{0}^{A(R)} = \left(\pi^0 - \hat{h}_0 \mp i\eta\right)^{-1}$, where $\eta$ is a small positive real number and  $\mybf{\nabla_{\mybf{p}}}\left(\hat{G}_{0}^{{R(A)}}\right)^{-1} = -\hat{\mybf{v}}$. Using the eigenbasis of $\hat{h}_0$,  $\hat{h}_0|n\rangle=\epsilon_{n}|n\rangle$, to evaluate the trace and the residue theorem to perform the $\pi^0$ integration we find
	\begin{align}
		\tilde{\kappa}_{ij} & = \frac{1}{\hbar T} \sum_n \left(m_{n}^k \epsilon_{n} f_{\mp}(\epsilon_{n})-\Omega_{n}^k\epsilon_{n}^2 f_{\mp}(\epsilon_{n})\right), \label{E:tildek_ij2} \\
		M_Q^k & = \frac{1}{2\hbar} \sum_n \left(
		2 \Omega_{n}^k \int_{\infty}^{\epsilon_{n}} d\epsilon \epsilon f_{\mp}(\epsilon) -  m_{n}^k \epsilon_{n} f_{\mp}(\epsilon_{n})
		\right). \label{E:MQk}
	\end{align}
	The magnetic moment reads
	\begin{align} 
		m_{n}^k &= -2\epsilon_{ijk}\text{Im} \sum_m \frac{ \langle n |\hat{v}_{i}| m \rangle \langle m |\hat{v}_{j}| n \rangle }{(\epsilon_{n}-\epsilon_{m})} \label{E:m}
	\end{align}
	and the Berry curvature reads
	\begin{align} 
		\Omega_{n}^k &= -2\epsilon_{ijk}\text{Im} \sum_m \frac{ \langle n |\hat{v}_{i}| m \rangle \langle m |\hat{v}_{j}| n \rangle }{(\epsilon_{n}-\epsilon_{m})^2}. \label{E:Omega}
	\end{align}
	By combining Eqs. (\ref{E:tildek_ij2}) and (\ref{E:MQk}) we arrive at
	\begin{align} \label{E:k_ij2}
		{\kappa}_{ij} & = -\frac{1}{\hbar T} \sum_n \Omega_{n}^k \left(\epsilon_{n}^2 f_{\mp}(\epsilon_{n}) -2 \int_{\infty}^{\epsilon_{n}} d\epsilon \epsilon f_{\mp}(\epsilon) \right) \nonumber\\
		&= - \frac{1}{\hbar \beta^2 T} \sum_n \Omega_{n}^k c_2(f_{\mp}),
	\end{align}
	where 
	\begin{align} \label{E:c2}
		c_2(f_{\mp}) = - \int^{\infty}_{\epsilon_{n}} d\epsilon
		(\beta\epsilon)^2 f_{\mp}'(\epsilon).
	\end{align}
	This is the expression derived in Refs.\ \cite{PhysRevLett.106.197202, Matsumoto2011, Murakami2011, PhysRevB.84.184406, PhysRevB.89.054420,PhysRevLett.107.236601,Shitade2014, Murakami2017}.
	
	\begin{figure}[b]
		\includegraphics[width=\linewidth]{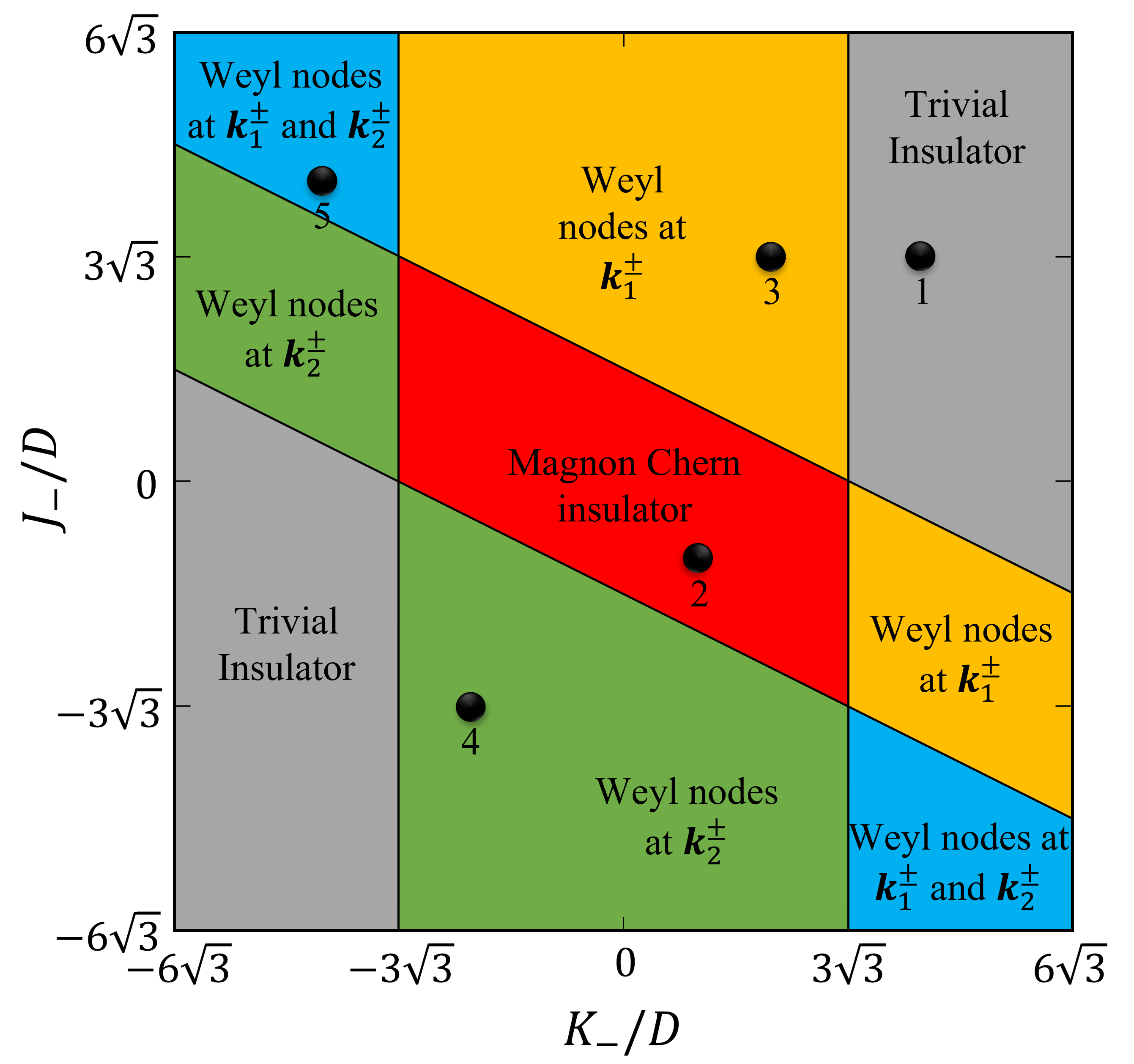}
		\caption{Phase diagram of the stacked honeycomb ferromagnet in the $(K_-,J_-)$ plane. The black dots mark the points chosen to analyze the behavior of the Chern number and thermal conductivity in phases 1 to 5.}
		\label{fig:phase_diagram}
	\end{figure}
	
	\begin{figure*}
		\includegraphics[width=\linewidth]{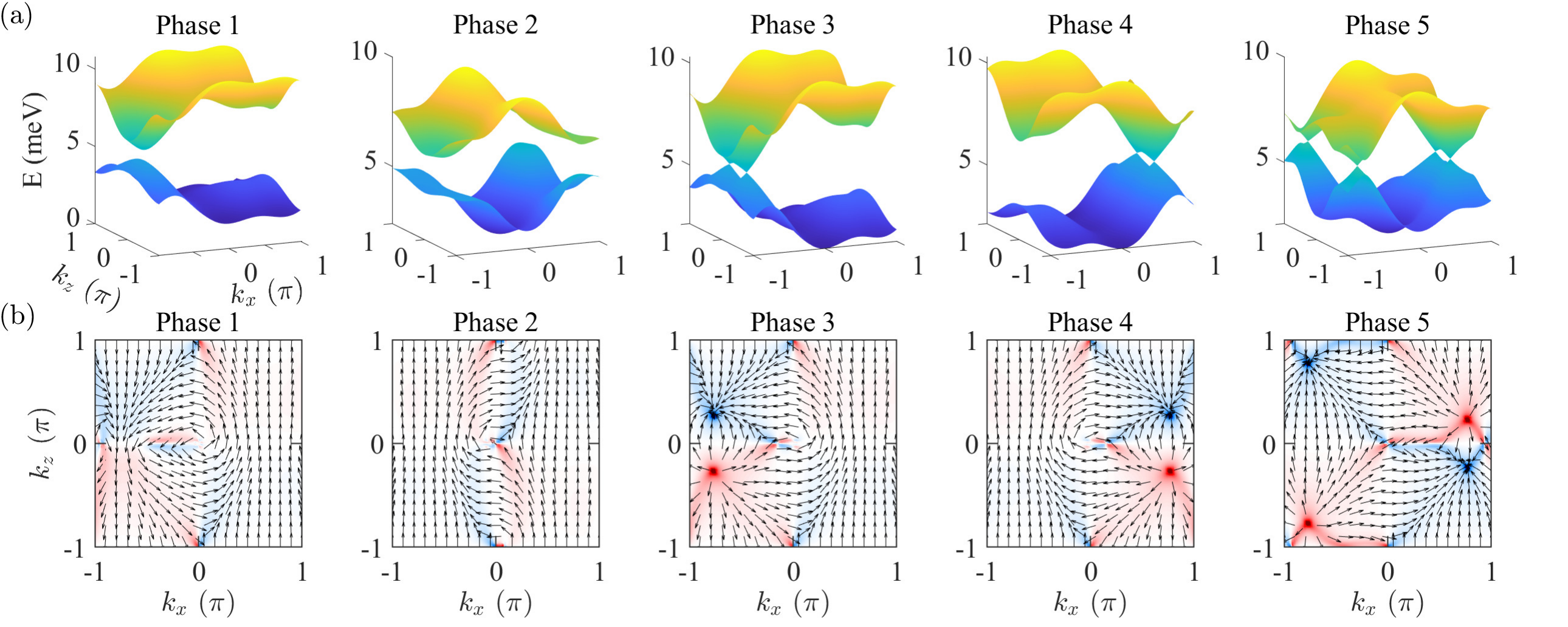}
		\caption{(a) Band structure and (b) Berry curvature of the lower magnon band in the $k_xk_z$ plane at $k_y=0$. WM nodes are evident in phases 3 to 5. The Berry curvature vector is indicated by arrows. The color represents the divergence of the Berry curvature with blue indicating negative values and red indicating positive values.}
		\label{fig:BS_BC}
	\end{figure*}
	
	\begin{figure*}[t]
		\includegraphics[width=\linewidth]{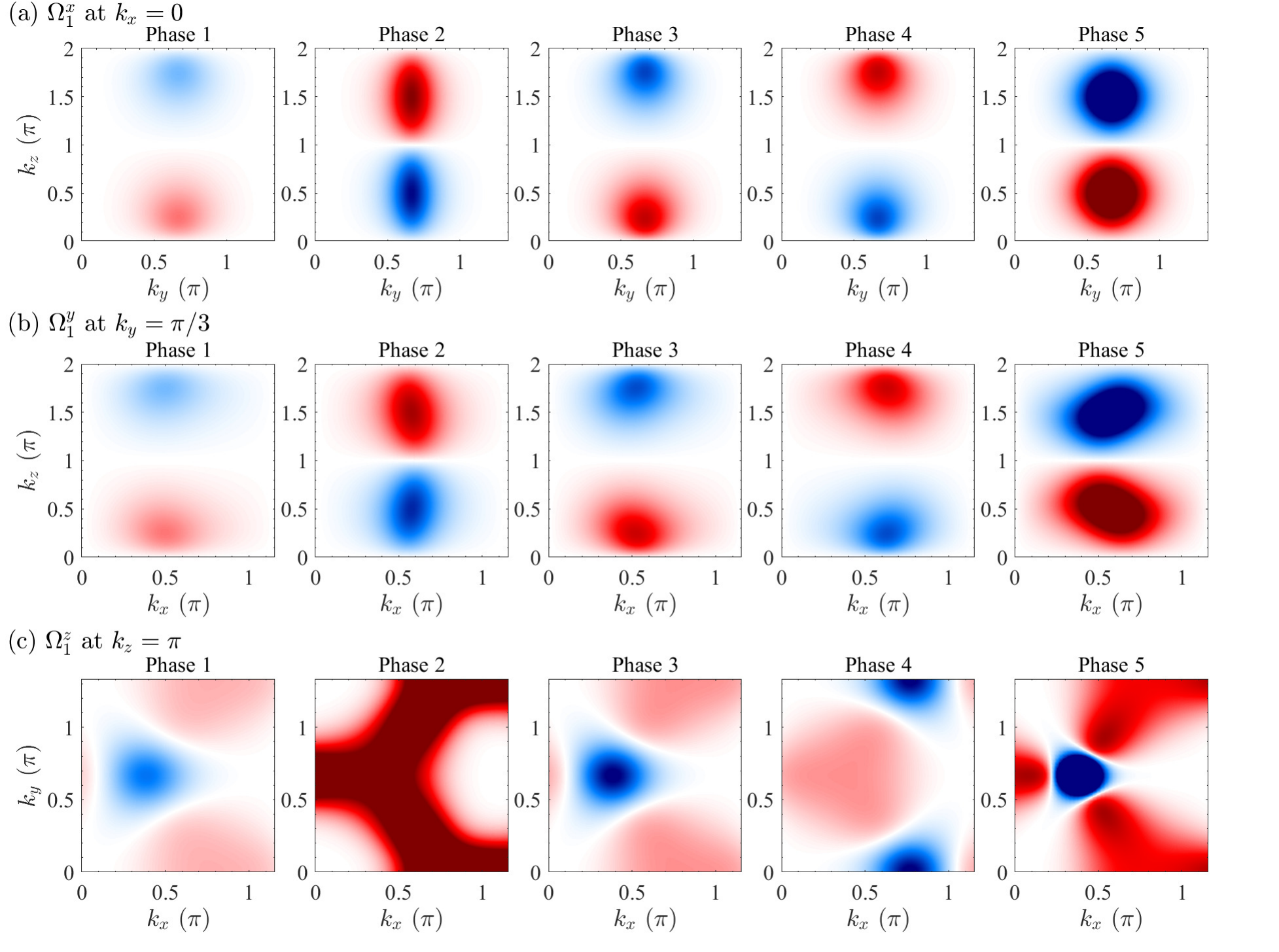}
		\caption{Berry curvature of the lower magnon band: (a) $\Omega_{1}^x$ at $k_x=0$, (b) $\Omega_{1}^y$ at $k_y=\pi/3$, and (c) $\Omega_{1}^z$ at $k_z=\pi$. Blue color indicates negative values and red color indicates positive values.}
		\label{fig:BC_All}
	\end{figure*}
	
	\section{Thermal Hall effect of Weyl Magnons in a Stacked Honeycomb Ferromagnet} \label{sec:honeycomb_FM}
	\subsection{Model}
	The thermal Hall effects in two-dimensional ferromagnetic and antiferromagnetic magnonic systems (induced by spin-orbit coupling) were studied extensively \cite{PhysRevB.89.134409,PhysRevLett.104.066403,PhysRevLett.106.197202,PhysRevB.84.184406,PhysRevB.89.054420,PhysRevB.99.014427,PhysRevB.98.094419,PhysRevB.99.054409,PhysRevB.95.014422,PhysRevLett.121.097203,PhysRevB.94.174444,PhysRevLett.122.057204}, while less attention was given to three-dimensional magnonic systems. In analogy to electrons, magnons can exhibit non-trivial topological properties. One example are the Weyl magnons (WMs) in three-dimensional magnets \cite{PhysRevB.97.094412,Owerre2018,PhysRevLett.117.157204}. To study the impact of WMs on $\kappa_{ij}$, we address the toy model of the stacked honeycomb ferromagnet shown in Fig.\ \ref{fig:honeycomb}, whose topological properties were determined in Ref.\ \cite{PhysRevB.96.104437}. The Hamiltonian reads
	\begin{align} \label{E:H_honeycomb}
		H = & -J\sum_{\langle i,j \rangle,n} \mybf{S}_{i,n} \cdot \mybf{S}_{j,n} + \sum_{\llangle  i,j  \rrangle } \mybf{D}\cdot \left(\mybf{S}_{i,n}\times \mybf{S}_{j,n}\right) \\ \nonumber
		&-\sum_{i,n} K_i \left(S_{i,n}^z\right)^2 -\sum_{i,\langle n,m \rangle } J_i \mybf{S}_{i,n} \cdot \mybf{S}_{i,m},
	\end{align} 
	where $\langle \cdots \rangle $ and $\llangle \cdots \rrangle $ represent the nearest neighbor and next-nearest neighbor sites, respectively, and $\mybf{S}_{i,n}$ represents the spin at site $i$ (A or B) of layer $n$. The first term is the symmetric Heisenberg interaction with $J>0$ and the second term is the antisymmetric Dzyaloshinskii-Moriya interaction with Dzyaloshinskii-Moriya vector $\mybf{D}$. Only the perpendicular component $D \nu_{ij}$ (with $\nu_{ij}$ alternating between $-1$ and $1$ on consecutive intralayer bonds) of $\mybf{D}$ survives \cite{Dzyaloshinsky1958,PhysRev.120.91}. The third term is the anisotropy energy and the last term is the interlayer interaction. A slightly different model with third-nearest neighbors was proposed in Ref.\ \cite{PhysRevB.104.104419}. However, this model has only either WM phases or magnon Chern insulator phases with bands of Chern number either $-2$ or $1$, while our aim is to compare $\kappa_{ij}$ of different phases, including a trivial insulator phase with bands of Chern number $0$.
	
	We start our analysis of $H$ with the standard linear spin-wave method \cite{Petit2011, Toth2015}. Focusing solely on the lowest order in the magnon creation ($\gamma_{i,n}^\dagger$) and annihilation ($\gamma_{i,n}$) operators, we utilize the Holstein-Primakoff transformation \cite{PhysRev.58.1098} 
	\begin{align} \label{E:HP_transformation}
		\begin{aligned}
			S_{i,n}^{+}=&S_{i,n}^x + iS_{i,n}^y=\sqrt{2S_{i,n}}\sqrt{1-{\frac{\gamma_{i,n}^{\dagger}\gamma_{i,n}}{2S_{i,n}}}} \gamma_{i,n},  \\
			S_{i,n}^{-}=&S_{i,n}^x - iS_{i,n}^y=\sqrt{2S_{i,n}}\gamma_{i,n}^{\dagger}\;\sqrt{1-{\frac{\gamma_{i,n}^{\dagger}\gamma_{i,n}}{2S_{i,n}}}}\;,\\
			S_{i,n}^{z}=&(S_{i,n}-\gamma_{i,n}^{\dagger}\gamma_{i,n}).
		\end{aligned}
	\end{align}
	Keeping only the quadratic terms, we then perform a Fourier transformation using the relation
	\begin{equation} \label{E:magnon_FT}
		\gamma_{i,n}^{(\dagger)}=\frac{1}{\sqrt{L}}\sum_{\mybf{k}}\text{e}^{(-)i\mybf{r}_{i,n}\cdot \mybf{k}}\gamma_{\mybf{k}}^{(\dagger)},
	\end{equation}	
	where $L$ denotes the number of unit cells, $\gamma_{\mybf{k},n}^{\dagger}$ and $\gamma_{\mybf{k},n}$ denote the magnon operators in momentum space, and $\mybf{r}_{i,n}$ denotes the position vector. We obtain 
	\begin{equation}
		H = \sum_{\mybf{k}} \Gamma_{\mybf{k}}^\dagger h(\mybf{k}) \Gamma_{\mybf{k}}
	\end{equation}
	with the basis vectors $\Gamma_{\mybf{k}}$ and 
	\begin{equation}
		h(\mybf{k}) = \sum_{\alpha\;=\; 0,x,y,z} h_\alpha(\mybf{k}) \sigma_\alpha,
	\end{equation}
	where $\sigma_0$ is the identity matrix, $\sigma_{x}$, $\sigma_{y}$, and $\sigma_{z}$ are the Pauli matrices, and
	\begin{align}
		\begin{split}
			h_0(\mybf{k}) & = 3JS + K_+S + J_+S\left(1-\cos(k_z c)\right), \\
			h_x(\mybf{k}) & = -JS\sum_{\beta=1}^3 \cos(\mybf{k}\cdot \mybf{d}_\beta^{(1)}), \\
			h_y(\mybf{k}) & = -JS\sum_{\beta=1}^3 \sin(\mybf{k}\cdot \mybf{d}_\beta^{(1)}), \\
			h_x(\mybf{k}) & = 2DS\sum_{\beta=1}^3 \sin(\mybf{k}\cdot \mybf{d}_\beta^{(2)}) + K_-S+J_-S\left(1-\cos(k_z c)\right)
		\end{split}
	\end{align}
	with $K_{\pm}=K_\text{A}\pm K_\text{B}$ and $J_{\pm}=J_\text{A}\pm J_\text{B}$. The vectors $\mybf{d}_\beta^{(1)}$ between nearest neighbor sites and $\mybf{d}_\beta^{(2)}$ between next-nearest neighbor sites are shown in Fig.\ \ref{fig:honeycomb}.
	
	\begin{figure*}
		\includegraphics[width=\linewidth]{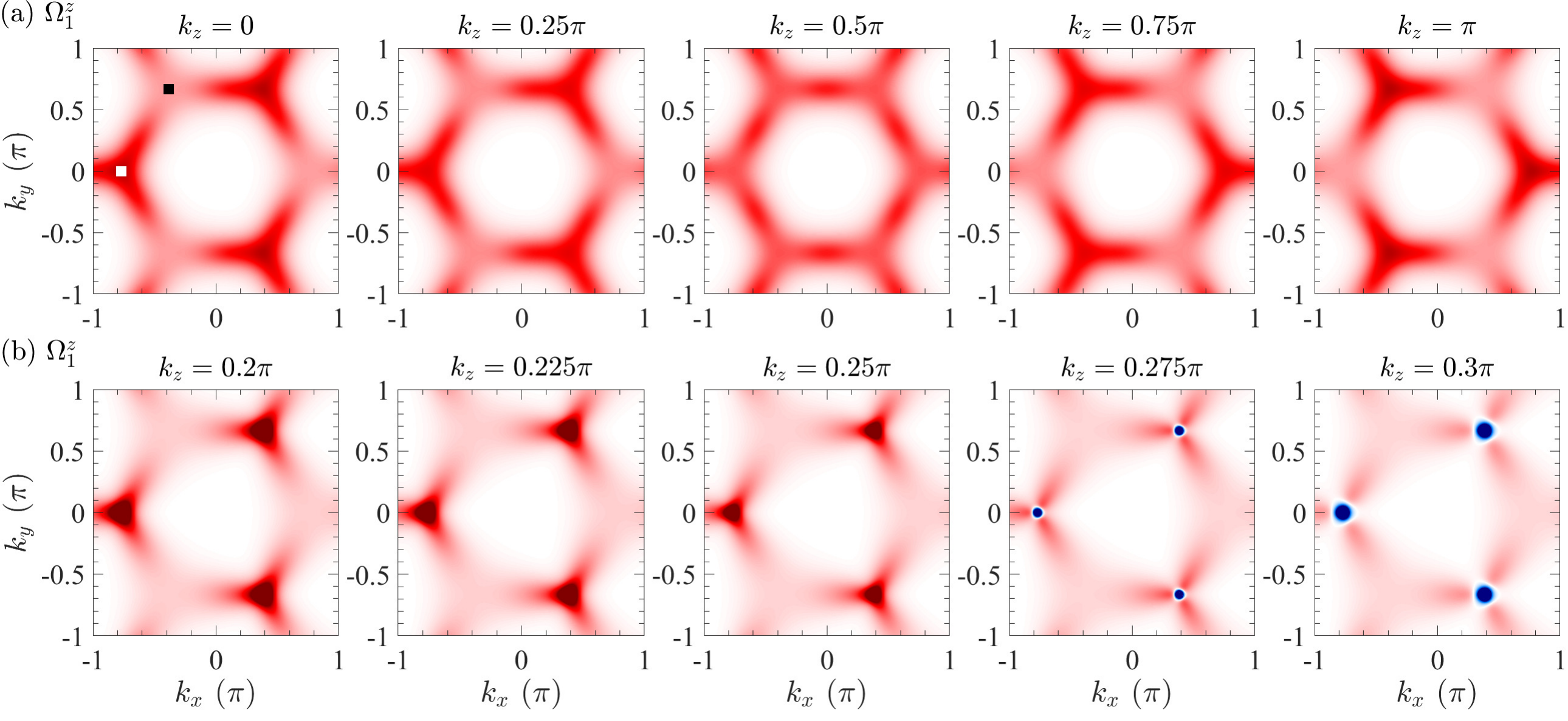}
		\caption{Berry curvature ($z$ component) of the lower magnon band for (a) phase 2 and (b) phase 3 at different values of $k_z$. The maxima transfer from the K point (white square) to the K' point (black square) as $k_z$ changes from $0$ to $\pi$. In phase 3 the maxima become minima abruptly as $k_z$ passes through the WM node at $k_z=\cos^{-1}(K_-/J_- + 1 - 3\sqrt{3}D/J_-)\approx 0.2677\pi$. Blue color indicates negative values and red color indicates positive values.}
		\label{fig:BC_vs_kz}
	\end{figure*}

	The phase diagram in the $(K_-,J_-)$ plane is constructed for $D=0.2J, K_+ = 12D$, and $J_+=2D$ (following Ref.\ \cite{PhysRevB.96.104437}). The phase boundaries are obtained by evaluating $h_{x}(\mybf{k})=h_{y}(\mybf{k})=h_{z}(\mybf{k})=0$ to determine the parameters at which bands cross each other. This gives rise to four boundaries, $K_- = \pm 3\sqrt{3} D$ and $K_-=-2J_-\pm 3\sqrt{3}D$, and generates one gapped trivial insulator phase (phase 1), one gapped magnon Chern insulator phase (phase 2), and three gapless WM phases (phases 3 to 5), as shown in Fig.\ \ref{fig:phase_diagram}. Phase 3 has a pair of WM nodes at $\left(-4\pi/3\sqrt{3}, 0, \pm \cos^{-1}(K_-/J_- + 1 - 3\sqrt{3}D/J_-)\right)$, phase 4 has a pair of WM nodes at $\left(-4\pi/3\sqrt{3}, 0, \pm \cos^{-1}(K_-/J_- + 1 + 3\sqrt{3}D/J_-)\right)$, and phase 5 combines the two pairs.
	
	To analyze the phases, we consider the points $(K_-,J_-)$ marked in Fig.\ \ref{fig:phase_diagram}: $(4\sqrt{3},3\sqrt{3})D$ for phase 1, $(\sqrt{3},-\sqrt{3})D$ for phase 2, $(2\sqrt{3},3\sqrt{3})D$ for phase 3, $(-2\sqrt{3},-3\sqrt{3})D$ for phase 4, and $(-4\sqrt{3},4\sqrt{3})D$ for phase 5. The band structure and Berry curvature of the lower magnon band in the $k_xk_z$ plane at $k_y=0$ are addressed in Fig.\ \ref{fig:BS_BC}. Monopoles of the Berry curvature are present at the WM nodes in phases 3 to 5. 
	
	\begin{figure}
		\includegraphics[width=\linewidth]{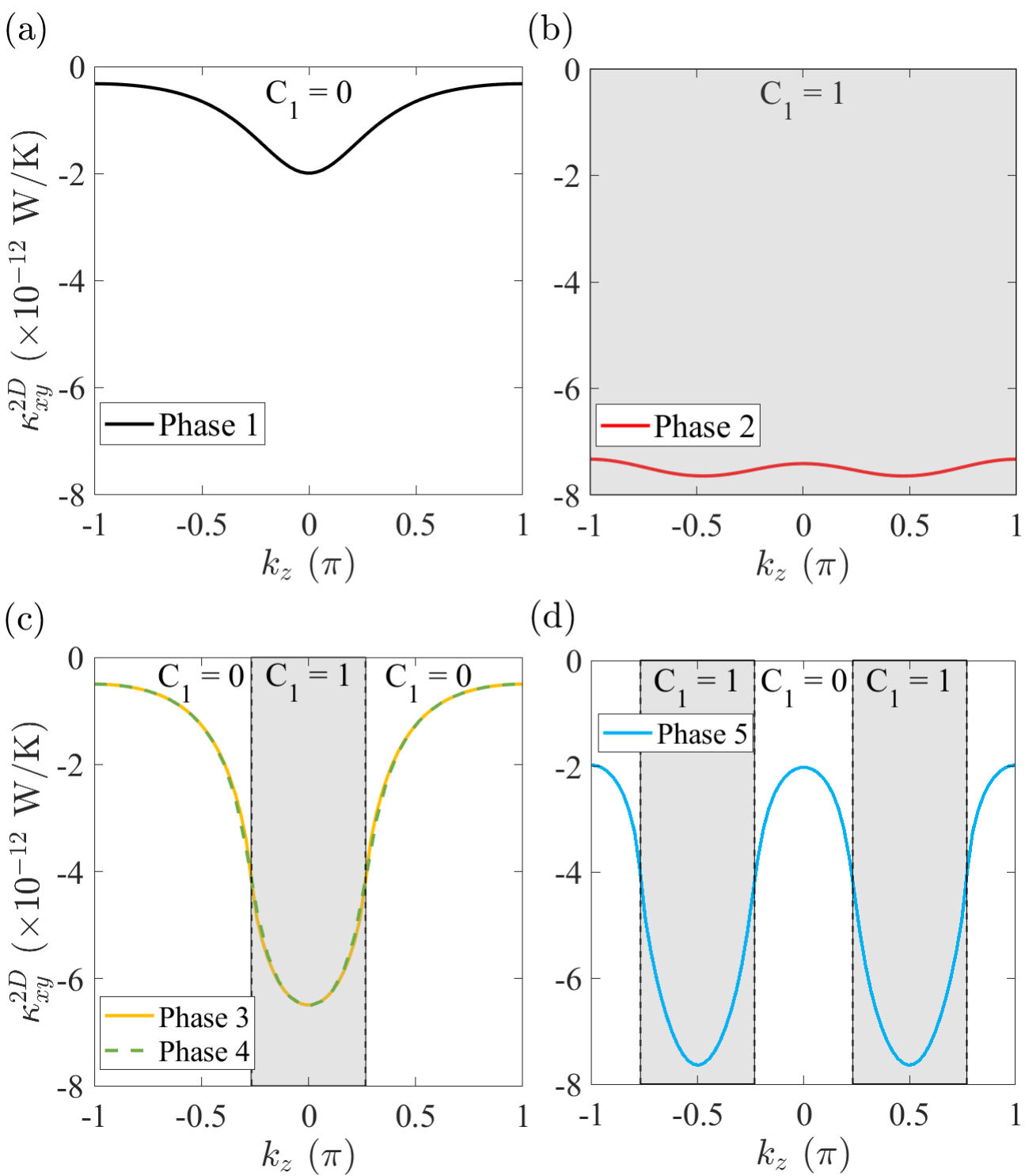}
		\caption{Two-dimensional thermal conductivity ($xy$ component) for (a) phase 1 (trivial insulator), (b) phase 2 (magnon Chern insulator), (c) WM phases 3 and 4, and (d) WM phase 5. The Chern number of the lower magnon band is indicated by white ($\text{C}_1$ = 0) and gray ($\text{C}_1$ = 1) colors.}
		\label{fig:chern_vs_kz}
	\end{figure}
	
	\subsection{Chern number and thermal conductivity}
	We decompose the three-dimensional Hamiltonian into two-dimensional Hamiltonians indexed by $k_z$,
	\begin{equation}
		h(\mybf{k})=\sum_{k_z}h_{k_x,k_y}(k_z),
	\end{equation}
	and calculate the Chern number of the $n$th magnon band ($n=1$ or $2$) in the traditional way as
	\begin{equation} \label{E:chern}
		\text{C}_n(k_z) = \frac{1}{2\pi} \int dk_y \int dk_x \Omega_{n}^z(k_z).
	\end{equation}
	Similarly, Eq.\ (\ref{E:k_ij2}) implies for the three-dimensional thermal conductivity
	\begin{equation} \label{E:kappa3D}
		\kappa_{ij}^{3D} = \int \frac{dk_z}{2\pi} \kappa_{ij}^{2D}(k_z),
	\end{equation}
	where
	\begin{equation} \label{E:kappa2D}
		\kappa_{ij}^{2D}(k_z) = -\frac{1}{\hbar \beta^2 T}\sum_n \int dk_y \int dk_x \Omega_{n}^k(k_z)c_2(f_+)
	\end{equation}
	is the two-dimensional thermal conductivity.
	
	\begin{figure*}
		\includegraphics[width=\linewidth]{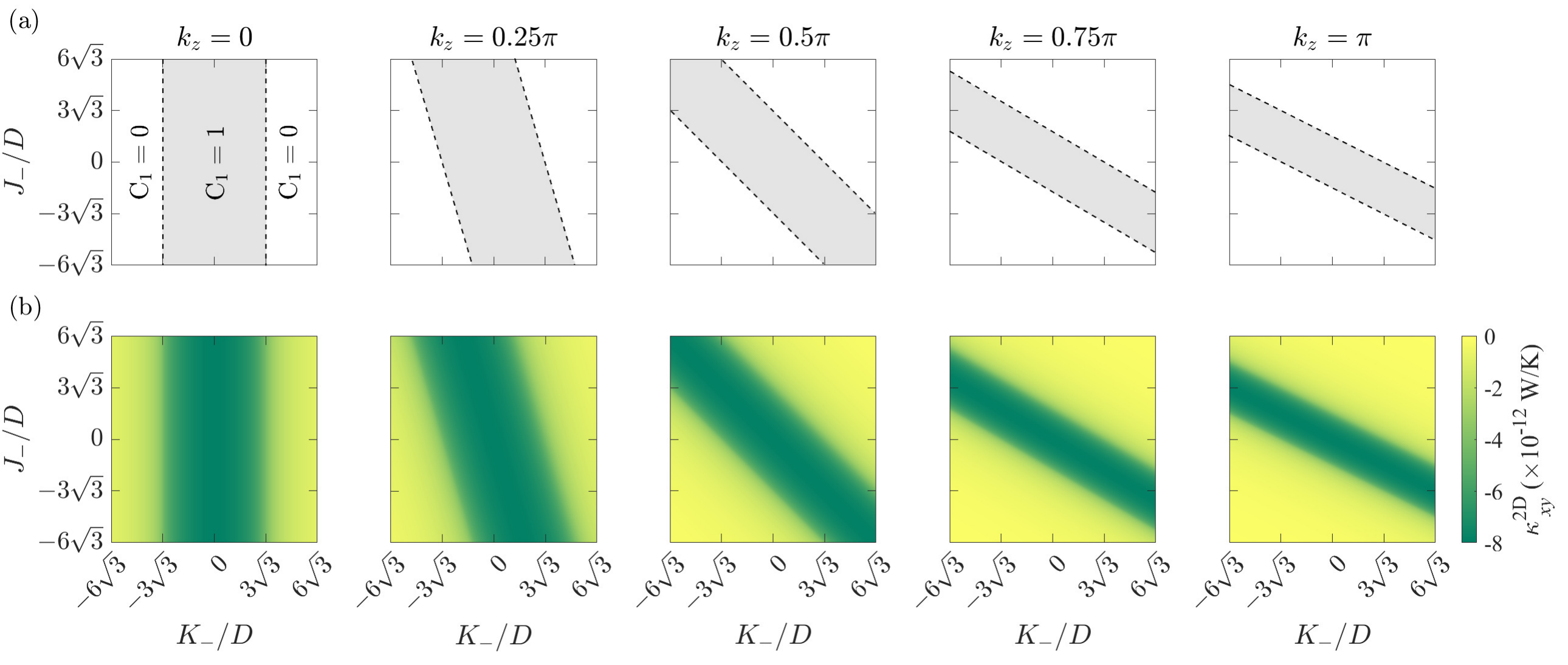}
		\caption{(a) Chern number of the lower magnon band and (b) two-dimensional thermal conductivity ($xy$ component) at $T$ = 100 K in the ($K_-, J_-$) plane for different values of $k_z$.}
		\label{fig:Chern_and_cond_in_KmJm}
	\end{figure*}
	
	WM nodes appear in pairs with opposite chirality for planes not orthogonal to the $k_z$ axis, resulting in zero net Berry curvature, as illustrated in Fig.\ \ref{fig:BC_All}(a,b) for the $k_yk_z$ and $k_xk_z$ planes. As a consequence, $\kappa_{yz}^{3D}$ and $\kappa_{zx}^{3D}$ are zero. On the other hand, for planes perpendicular to the $k_z$ axis, see Fig.\ \ref{fig:BC_All}(c) for the $k_xk_y$ plane, the net Berry curvature is non-zero, leading to non-zero $\kappa_{xy}^{3D}$. Focusing on the $k_xk_y$ plane, we examine how the $z$ component of the Berry curvature, $\Omega_{n}^z$, behaves at different values of $k_z$ to understand the contribution of the WM nodes to $\kappa_{xy}^{3D}$. Note that in the WM phases most contributions are due to the lower magnon band ($n=1$) because of the Bose distribution. It turns out that $\Omega_{1}^z$ is minimal at $k_z=\pm \pi$ and maximal at $k_z=0$ in phase 1. It is an even function of $k_z$ in phase 2 with maxima that are roughly ten times higher than in phase 1. Moreover, in phase 2 the maxima transfer from the K point to the K' point as $k_z$ changes from $0$ to $\pi$, see Fig.\ \ref{fig:BC_vs_kz}(a). In all the WM phases the maxima of $\Omega_{1}^z$ become minima abruptly as $k_z$ passes through a WM node, see Fig.\ \ref{fig:BC_vs_kz}(b) for the WM node at $k_z = \cos^{-1}(K_-/J_- + 1 - 3\sqrt{3}D/J_-)\approx 0.2677\pi$ in phase 3. This results in a smaller $\Omega_{1}^z$ than that of phase 2, leading to a smaller $\kappa_{xy}^{3D}$. 

	Figure \ref{fig:chern_vs_kz} presents $\text{C}_\text{1}$ and $\kappa_{xy}^{2D}$ as functions of $k_z$ for all five phases. Phase 1 (trivial insulator) has $\text{C}_\text{1}=0$ and phase 2 (magnon Chern insulator) has $\text{C}_\text{1}=1$. $\text{C}_\text{1}$ of the WM phases changes between 0 and 1 as $k_z$ passes through WM nodes. $\kappa_{xy}^{2D}$ remains almost constant in phases 1 and 2, being significantly larger for the magnon Chern insulator. While it is influenced significantly by $\text{C}_\text{1}$ in the WM phases. Phases 3 and 4 show the same values because of identical $|K_-|$ and $|J_-|$. Figure \ref{fig:Chern_and_cond_in_KmJm} presents $\text{C}_\text{1}$ and $\kappa_{xy}^{2D}$ at $T$ = 100 K in the ($K_-, J_-$) plane. $\kappa_{xy}^{2D}$ transitions a gradually near the topological boundaries, which can be attributed to the emergence of chiral edge states, a characteristic feature of two-dimensional magnon Chern insulators. As $T$ increases, resulting in a higher magnon density, the transition at the topological boundaries becomes more pronounced.
	
\begin{figure}
		\includegraphics[width=\linewidth]{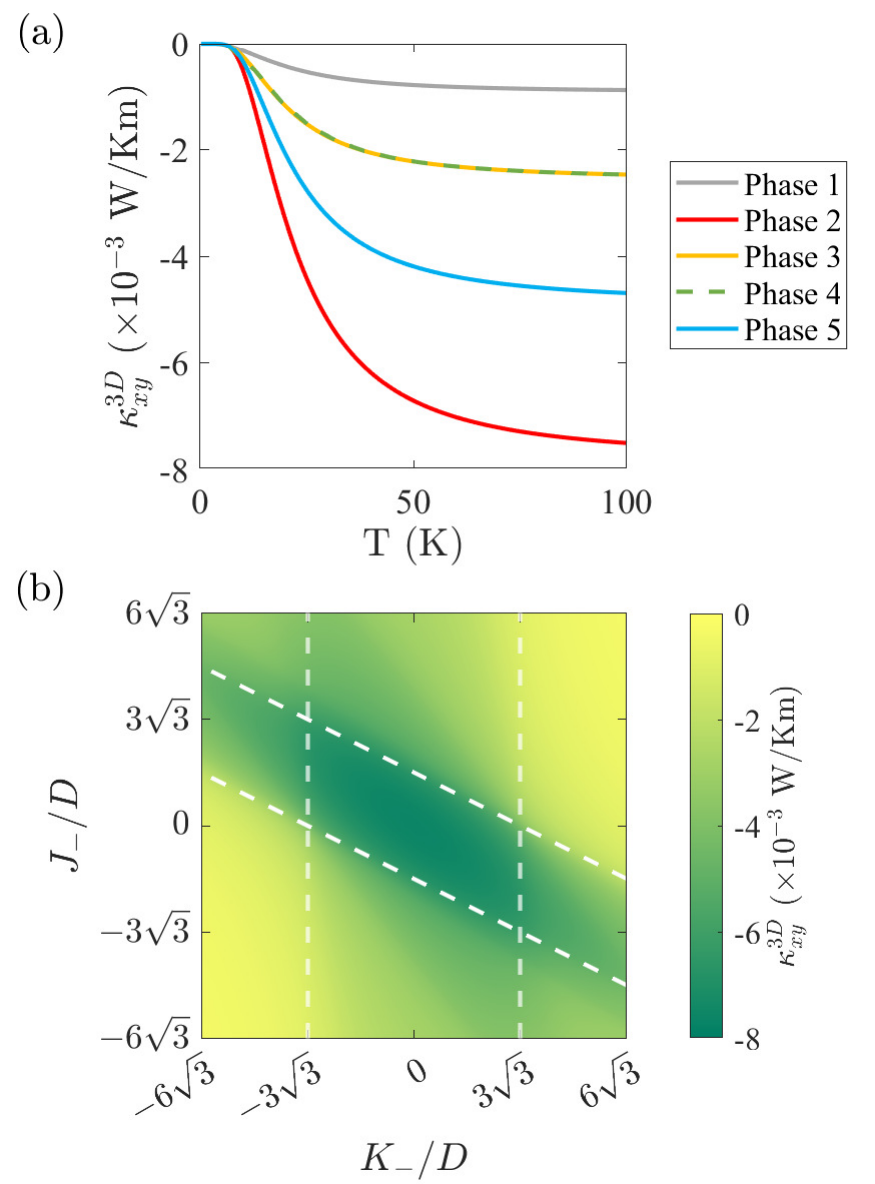}
		\caption{Three-dimensional thermal conductivity ($xy$ component) (a) as a function of $T$ and (b) at $T$ = 100 K in the ($K_-,J_-$) plane.}
		\label{fig:kappa3D_vs_T}
	\end{figure}

	Figure \ref{fig:kappa3D_vs_T} presents $\kappa_{xy}^{3D}$ as a function of $T$ and at $T$ = 100 K in the ($K_-, J_-$) plane. The absolute value increases for increasing $T$ in all the phases.	The highest absolute values are observed for the magnon Chern insulator, which is not surprising, because in this phase, unlike the WM phases, $\Omega_1^z$ maintains its sign throughout the $k_xk_y$ plane, as shown in Fig.\ \ref{fig:BC_vs_kz}(a).

	\section{Summary and Discussion} \label{sec:summary}
	We presented a theoretical framework for investigating the response of heat currents to a temperature gradient using the Keldysh-Dyson formalism. The core of this approach is an expansion of the Dyson equations through the Moyal product within the Wigner representation. The temperature gradient is described using both Luttinger's scalar potential ($\phi$) and Moreno-Coleman-Tatara's vector potential ($\mybf{A}_T$). While the thermal field does not arise from a formal gauge symmetry, we can achieve ``gauge invariance" under the transformations $\phi\to\phi+\frac{d\chi}{dt}$ and $\mybf{A}_T\to\mybf{A}_T-\mybf{\nabla}\chi$ by assigning a portion of the temperature gradient to $\phi$ and the remainder to $\mybf{A}_T$. This establishes a one-to-one correspondence between the responses induced by an electromagnetic field and by a temperature gradient.
	
	We derived the thermal conductivity tensor encompassing contributions from the heat current and heat magnetization, applicable to general multiband Hamiltonians. Our formalism treats the heat current and heat magnetization on equal footing, mirroring the usage of the method of Ref.\ \cite{Onoda2006} for calculating the electric current density and subsequently calculating the orbital magnetization \cite{PhysRevB.86.214415}. Our result for clean limit reduces to the well established expression \cite{PhysRevLett.106.197202, Matsumoto2011, Murakami2011, PhysRevB.84.184406, PhysRevB.89.054420, Murakami2017,Shitade2014,PhysRevLett.107.236601}. 
	\noindent The introduced approach has several advantages over previous formalisms. Firstly, it readily accounts for diamagnetic currents by replacing the canonical momentum with the kinetic momentum. Secondly, it can be applied to interacting and/or disordered systems. Thirdly, its ability to address non-linear responses extends beyond the scope of formalisms based on the Kubo formula. Fourthly, it can be extended to other transport quantities (e.g., particle transport, nonequilibrium densities) and time-dependent non-uniform problems.
	
	Using the derived formula for the thermal conductivity in the clean limit, the thermal properties of a stacked honeycomb ferromagnet were analyzed for all its phases (trivial insulator, magnon Chern insulator, and three WM phases with WM nodes at different points in the momentum space). Decomposing the three-dimensional Hamiltonian into two-dimensional Hamiltonians indexed by $k_z$, we observed that the Chern number remains constant for all values of $k_z$ for the trivial and magnon Chern insulator phases, while it changes abruptly from 0 to 1 as $k_z$ passes through WM nodes. Accordingly, in the trivial and magnon Chern insulator phases the $xy$-component of the two-dimensional thermal conductivity exhibits only minor variations as a function of $k_z$, while it changes significantly as $k_z$ passes through WM nodes in the WM phases. The primary contributions to the three-dimensional thermal conductivity in the WM phases are due to the WM nodes (monopoles of the Berry curvature). While one might expect the absolute value of the $xy$-component of the three-dimensional thermal conductivity to be higher in the WM phases than in the other two phases, interestingly, the magnon Chern insulator exhibits the highest absolute values, which can be understood by noting that in the WM phases the Berry curvature changes sign as $k_z$ passes through WM nodes.

	\begin{acknowledgments}
		The authors acknowledge insightful discussions with Gen Tatara, Konstantinos Sourounis, and Diego Garcia Ovalle. This work was supported by King Abdullah University of Science and Technology.
	\end{acknowledgments}

	\bibliography{bibolography}
	
\end{document}